\documentclass[12pt]{article}
\usepackage{latexsym}
\usepackage{amsmath}
\usepackage{amssymb}
\usepackage{tabularx}
\usepackage[ansinew]{inputenc} 
 \usepackage{pgfplots}
 \usepackage{tikz}
 \usetikzlibrary{plotmarks} 
\newtheorem{Theorem}{Theorem}[section]
\newtheorem{Definition}[Theorem]{Definition}

\newtheorem{Lemma}[Theorem]{Lemma}

\begin{document}
\sloppy

\begin{center} 
{\Large \bf Computation of the ``Enrichment'' of a Value 
Functions \\ of an Optimization Problem  on Cumulated 
Transaction-Costs\\ through a Generalized Lax-Hopf Formula}\\   
\mbox{}\\ 
 \emph{\textbf{Chen 
Luxi}}\footnote{Université Panthéon-Sorbonne and Société VIMADES 
(Viabilité, Marchés, Automatique et Décision), e-mail: 
clxshd@gmail.com}  
\\\today
\end{center}

\subsubsection*{Abstract}
The \emph{Lax-Hopf formula  simplifies the value function  of an 
intertemporal optimization (infinite dimensional) problem 
associated with a convex \emph{transaction-cost} function which 
depends only on the transactions (velocities) of a commodity 
evolution: it states that \emph{the value function is equal to 
the marginal fonction of a finite dimensional problem with 
respect to durations and average transactions},  much simpler to 
solve. The average velocity of the value function on a investment 
temporal window is regarded as an \emph{enrichment}, proportional 
to the profit and inversely proportional to the investment 
duration.\\ At optimum, the Lax-Hopf formula implies that \emph{the 
enrichment is equal to the cost  of the  average transaction on the investment temporal window.} \\
In this study, we generalize the Lax-Hopf formula when the 
transaction-cost function \emph{depends also on time and 
commodity}, for reducing the infinite dimensional problem to a 
finite dimensional problem. For that purpose, we introduce the 
\emph{moderated transaction-cost} function which depends only on 
the duration and on a commodity. \\Here again, the generalized 
Lax-Hopf formula reduces the computation of the value function to 
the marginal fonction of an optimization problem on durations and 
commodities involving the moderated transaction cost function. At 
optimum, the enrichment of the value function is still equal to 
the moderated transition cost-function of average transaction.}

\subsubsection*{Mathematics Subject Classification:} 
34A60, 90B10, 90B20, 90B99, 93C10, 93C30, 93C99 
 
\subsubsection*{Keywords:} Lax-Hopf formula, transaction of commodities, 
transaction-cost function, value-fonction of an intertemporal 
optimization problem,  generalized Lax-Hopf formula, moderation 
of a transaction cost-function,   terminal conditions

\section{Introduction}

Given a value function $t \mapsto V(t) \in \mathbb{R}^{}$ (a 
cash-flow, for instance, or the value function of an 
intertemporal optimal control problem), we interpret  its 
\emph{average velocity} $ \displaystyle{ 
\frac{V(T)-V(T-\Omega)}{\Omega}}$  on the temporal window 
$T-\Omega ,T]$ as its \index{enrichment} 
\emph{enrichment}\footnote{Its \index{(forward) interest rate} 
\emph{(forward) interest rate} is  
$\displaystyle{\frac{V(T)-V(T-\Omega )}{\Omega V(T-\Omega )} }$, 
its \index{(backward) interest rate} \emph{(backward) interest 
rate} is $\displaystyle{\frac{V(T)-V(T-\Omega )}{\Omega V(T)} }$ 
and its \index{(symmetric) interest rate} \emph{(symmetric) 
interest rate} $\displaystyle{\frac{V(T)-V(T-\Omega )}{\Omega 
\sqrt[]{V(T)V(T-\Omega )}}}$ of this investment. They all 
converge to the (instantaneous) interest rate  
$\displaystyle{\frac{V'(t)}{V(t)}}$ when $\Omega \rightarrow 0+ $ 
when $V(\cdot)$ is differentiable from the left.}, where $\Omega  
\geq 0$ is the aperture\footnote{The inverse  
$\displaystyle{\frac{1}{\Omega }}$ of the aperture $\Omega $ of a 
temporal window can be regarded as a definition of the notion of 
\emph{``liquidity''} (or \emph{``velocity''}, as is also called, 
although it does not mention   the variable of which  the inverse 
of aperture is the velocity in mathematical terminology). So, the 
enrichment is the product of the liquidity and the profit.} (or 
duration) of the temporal window.   

If we interpret $V(T)-V(T-\Omega )$ as a \emph{profit}, then the 
enrichment  $ \frac{V(T)-V(T-\Omega )}{\Omega }$ is \emph{the 
ratio of the  profit over the aperture of the temporal window}. 
The larger the profit, the smaller the aperture, the larger the 
ennrichment\footnote{This ratio could be a basis for a 
\emph{``Shareholder Value Tax''}  inversely proportional to the 
investment duration $\Omega$ and proportional to the profit   
(see Section~1.4, p. 18, of \emph{Time and Money. How Long and 
How Much Money is Needed to Regulate  a Viable  
Economy},\cite[Aubin]{TM}).}.

Let us introduce $X:=\mathbb{R}^{\ell}$,  regarded as a 
\index{commodity space} \emph{commodity space} of commodities 
$x:=(x_{h})_{1 \leq h \leq \ell}$ of amounts $x_{h} \in 
\mathbb{R}^{}$ of units $e^{h}$ of   goods or services labelled 
$h = 1, \ldots,\ell$ (see an economic motivation based on a 
dynamical version of the Willingness to Pay issue in 
Section~\ref{WTP}, p. \pageref{WTP}, leading to such an 
intertemporal optimization problem, among many other economic 
examples and issues). The velocity $x'(t)$ at time $t$  of the 
evolution of a commodity $x(\cdot)$ is regarded as  a 
\index{transaction} \emph{transaction} (actually, an 
infinitesimal one) since it is the limit of average transactions 
$\displaystyle{\frac{x(T)-x(T-\Omega )}{\Omega }}$ on the 
temporal window $[T-\Omega ,T]$ when the aperture converges to 
$0$\footnote{Derivatives from the left are used according to a 
suggestion of \glossary{Galperin (Efim A.) [1934-]} \emph{Efim 
Galperin}, since derivatives from the right $\displaystyle{x'(t)= 
\lim_{h \mapsto 0+}\frac{x(t+h)-x(t)}{h}}$ are ``physically 
non-existent''  since time $t+h$ is not yet known, following the  
 Jiri Buquoy, who in 1812, formulated the equation of 
motion of a body with variable mass, which retained only the 
attention of \glossary{Poisson (Siméon Denis) [1781-1840]} 
\emph{Poisson} before being forgotten. This is also the reason 
why we use temporal windows $[T-\Omega ,T]$ where $T$ is a flying 
present instead of using a present time $T \geq 0$ ranging over 
an unknown future, as it is currently done.}.

In this study, we shall take for value function the one provided 
by an intertemporal optimal control problem of the form 

 \begin{equation} \label{e:ValueFctChoux0} 
V(t,x) \;:= \; \inf_{\Omega \geq 0} \inf_{x(\cdot)  
}\left(\mathbf{c}(T-\Omega ,x(T-\Omega )) + \int_{T-\Omega 
}^{T}   \mathbf{l}(x'(t))dt\right) 
\end{equation} 
where the sum of the cost on the value $x(T-\Omega)$ at the 
beginning of the temporal window and the cumulated cost of the 
transactions   $\mathbf{l}(x'(t))$ on the temporal window are 
minimized  with respect to both the aperture $\Omega \geq 0$ and 
commodity evolutions defined below.

The question is to compute the enrichment 
$\displaystyle{\frac{V(T)-V(T-\Omega )}{\Omega} }$ in terms of 
the cost functions. When the transaction  is convex and  
continuous\footnote{Actually, lower semicontinuous.}, the 
celebrated  Lax-Hopf formula (see \cite[Hopf]{Hopf},  
\cite[Lax]{Lax}) states that at optimal aperture and optimal 
evolution, the enrichment is provided by the formula

\begin{equation} \label{e:}   
  \frac{V(T, 
x_{\star}(T))-V(T-\Omega_{\star}, 
x_{\star}(T-\Omega_{\star}))}{\Omega_{\star}} \; = \; 
\mathbf{l}\left( 
\frac{x_{\star}(T)-x_{\star}(T-\Omega_{\star})}{\Omega_{\star}}\right) 
 \end{equation}
stating that \emph{the cost of the average optimal transaction is 
the average velocity of the value function on the temporal 
window}, or, in economic terms, that  \emph{the enrichment is 
equal to the cost of the average transaction  of the optimal 
evolution}.

Once this formula recognized, we generalize this enrichment 
formula even \emph{when the transaction cost function depends on 
time and commodity} by proving a generalization of the Lax-Hopf 
formula in Theorem~\ref{t:GenLH}, p.\pageref{t:GenLH}. We next 
pass from the Willingness to Pay example to the case of an 
economy involving the evolution of both  commodities and their 
prices and taking for ``potential'' function the patrimonial 
value. Since its derivative involve not only the velocities of 
commodities (transactions) and the velocities of prices (price 
fluctuations, but also the values of the commodities and prices, 
the Lax-Hopf does not apply, but the generalized Lax-Hopf does.

\textbf{Organization of the Exposition:}

We begin by motivating the use of the Lax-Hopf by  an example 
intertemporal optimal control problem  derived from a dynamic 
version of the Willingness to Pay issue in Section~\ref{WTP}, p. 
\pageref{WTP}. Next, in Section~\ref{ValueFct}, p. 
\pageref{ValueFct}, we consider a version of an general optimal 
control problem posed on temporal windows $[T-\Omega ,T]$ of 
unknown aperture $\Omega \geq 0$ with terminal conditions instead 
of initial ones. When the transaction costs depend only on the 
transactions, we recall the Lax-Hop formula in Section~\ref{LHf}, 
p. \pageref{LHf}, generalized in Section~\ref{CLHf}, p. 
\pageref{CLHf} when the transaction costs depend also on time and 
commodities. In Section~\ref{IRCLHf}, p. \pageref{IRCLHf}, we 
take interest functions which are not fixed, but depend on time, 
commodities and transactions and extend to this case the 
generalized Lax-Hopf formula. We end this study in 
Section~\ref{DECLHf}, p. \pageref{DECLHf}, by applying the 
generalized Lax-Hopf formula to general economies involving 
commodities and prices.

\section{An Economic Motivation: Willingness to Pay} \label{WTP}

``Willingness To Pay''(resp. Accept) is defined in the literature 
as  \emph{the maximum amount a person would be willing to pay (in 
monetary units) of an exchange of an ``economic state" to receive 
(resp. accepting) the profit or avoid the sacrifice or something 
undesirable}, such as pollution. (See \cite[Hanemann]{Hanemann}, 
\emph{L'évaluation contingente : les valeurs ont-elles un prix 
?}, in \emph{Rendre possible,  Jacques Weber, itinéraire d’un 
économiste passe-frontières}, \cite[Bouamrane \emph{et 
al}]{Weber0} and \emph{Évaluation économique de la biodiversité} 
\cite[Brahic \& Terreaux]{BrahicTerreaux} de Brahic et J.-Ph. 
Terreaux and its bibliography,  among an  infinity of other 
publications on this topic).  Here, we follow the presentation of 
Chapter~6, p. 85, of \emph{Time and Money. How Long and How Much 
Money is Needed to Regulate  a Viable  Economy}, 
\cite[Aubin]{TM}).

A (static) economic perspective  of this concept requires a 
\emph{Willingness To Pay Valuation valuation (function)}, where 
$x \in \mathbb{R}^{\ell}$ is the ``economic state'' to evaluate 
and $w \in \mathbb{R}^{}$, the``wealth''. If $x_{0}$ and $ w_ {0} 
$  denote  the original state and its value, and $x$ and $w$ 
another state and its value, the question arises to compute the 
value $w$ of $x$ such that as \emph{$(x,w)$ has the same utility  
than $(x_{0},w_{0})$}, i.e., is a solution to 

\begin{equation} \label{e:WPEstatic}   
\mathbf{u}(x, w ) \; = \; \mathbf{u}(x_{0},w_{0})
\end{equation}

Then   $\varpi (x; x_{0}, w_ {0}):=w-w_{0}$ is the 
\emph{transaction cost} for obtaining $x$ from $x_{0}$, the 
\emph{``Willingness To Pay''} for exchanging $x_{0}$ with $x$  
defined implicitly as a solution to the equation 
\begin{equation} \label{e:ad3t-decisionRule}
\mathbf{u}(x, w_{0}+\varpi (x; x_{0}, w_ {0})) \; = \; 
\mathbf{u}(x_{0},w_{0})
   \end{equation} 
However,     transactions defined as instantaneous exchange 
$x'(t)$ of an evolving commodity $x(t)$, involve some underlying  
evolutionary (dynamical) process for exchanging  an initial 
commodity $x_{0}$ with a new one, which requires the introduction 
of 
\begin{enumerate}

\item a \emph{time} $ T \in \mathbb{R}^{}$ (evolving present 
time);

\item a \emph{duration} $ \Omega  \geq 0 $;

\end{enumerate} 
defining the \index{temporal window} \emph{temporal window} 
$[T-\Omega , T] $, the \emph{beginning} of which is $ T-\Omega $  
and the \emph{end}  of which is the present time $ T $, in  
which   the \emph{current past time} $t \in [T-\Omega ,T] $ 
evolves.

Instead of pairs of economic states $x \in X:=\mathbb{R}^{l}$ 
having the same ``Willingness To Pay Valuation'', one, $ x (T) $, 
at ending time $ T $, the successor of another one, $ x (T-\Omega 
) $, at the beginning of the temporal window.  In the example 
above, $(x,w )$ is regarded as the successor of $(x_{0}, w_{0})$. 
In this evolutionary context, this would mean that  the 
\emph{evolution between pairs  $ (x_{0}, w_ {0}) $ and $ (x, w) $ 
leaves constant the Willingness To Pay Valuation}.

In this dynamical framework, instead of assuming that a 
\emph{Willingness To Pay Valuation (function)} is  given, as in 
the static case, we shall built it from the following data: 

\begin{enumerate} 
\item an \index{evolutionary system} \emph{evolutionary system} 
governing a set $\mathcal{A}_{c}(T-\Omega ,T;x)$ of evolutions of 
the economic states $x(\cdot)$ defined on the temporal window 
$[T-\Omega ,T]$ and \index{arriving} \emph{arriving} at $x$ at 
time $T$ with velocities $x'(t)$ bounded by $c>0$: $\|x'(t) \| 
\leq c$.

  \item   an  ``(instantaneous) cost function'' $x 
\mapsto \mathbf{c}(T-\Omega ,x)$  indexed by the beginning of the 
temporal window, interpreted as the cost of an investment at the 
beginning of the temporal window.  
\end{enumerate}

This \emph{final condition} replaces the standard initial 
condition, since we are interested in \emph{time irreversible 
systems} when only the past (described on the temporal window 
$[T-\Omega , \Omega ] $)   is known and evolves with present time 
$T$. \emph{\emph{The value $x(T-\Omega )$ of the evolution 
$x(\cdot)$ at the beginning of the temporal  window is not 
prescribed}} in this study.  One can define  a  ``Willingness To 
Pay Valuation'' $(T,\Omega ;x) \mapsto V(T,\Omega ;x)$ in the 
following way: 
\begin{equation} \label{e:WTPfct0}
W(T,\Omega ;x) \; := \; \inf_{x(\cdot)\in 
\mathcal{A}_{c}(T-\Omega ,T;x)} \mathbf{c}(T-\Omega , x(T-\Omega 
))
\end{equation}
minimizing both the investment duration $\Omega \geq 0$ and the 
initial investment cost.  

We observe that by taking the zero duration $\Omega =0$, we 
derive from the construction of the Willingness To Pay valuation 
that the \emph{instantaneous boundary property} (for zero 
duration)
\begin{displaymath}   
\forall \; T, \; \forall \; x \in X, \; \; W(T,0;x) \; = \;  
\mathbf{c}(T, x)
\end{displaymath} 
holds true.   

One of the required properties for a function to be regarded as a 
``Willingness To Pay valuation function'' is that, as property~  
(\ref{e:WPEstatic}), p. \pageref{e:WPEstatic},  in the static 
case,  
\begin{displaymath}   
\forall \; t \in  [T-\Omega ,T], \; \;  W(t, t-(T-\Omega ), 
\overline{x}(t)) \; := \;  W(T,\Omega ;x)  
\end{displaymath}
(the dynamic programming property).

{\bf Remark} --- \hspace{ 2 mm}  When the economic state is a 
pair $(x,w) \in X \times \mathbb{R}^{}_{+} $ where $x \in X$ is a 
commodity and $w$ a wealth,  consider the instantaneous data 
$\mathbf{c}(t,x,w)$. The derived Willingness To Pay feedback map 
$(t,d,x,w)  \leadsto R (t,d,x,w)$ governs the Willingness To Pay 
evolutions according a differential inclusion
\begin{equation} \label{e:}   
\forall \; t \in [T-\Omega ,T], \; \; 
(\overline{x}'(t),\overline{w}'(t)) \; \in \; R(t,t-(T-\Omega 
),\overline{x}(t),  \overline{w}(t))
\end{equation}

Therefore, for any $t \in [T-\Omega , T]$,  
$\displaystyle{\overline{x}(T)-\overline{x}(t)=\int_{t}^{T}\overline{x}'(\tau)d\tau}$ 
is the \emph{transaction} between $t$ and $T$ and 
\begin{equation} \label{e:}   
\forall \;  t \in [T-\Omega , T], \; \varpi (x; x_{t}, w_ {t}) \; 
:= \; \int_{T-t}^{T} \overline{w}'(\tau) d\tau 
\end{equation}
is its \emph{transaction cost}, regarded as the Willingness To 
Pay in the static case which motivated this study. \hfill $\;\; 
\blacksquare$ \vspace{ 5 mm}

The ``average deprivation" in the case of a process  transforming 
$ x (T-\Omega ) $ at the beginning of the temporal window to $ x 
(T) $ at the end of this window is equal to  
$\displaystyle{\frac{x(T)-x(T-\Omega )}{\Omega } = 
\frac{1}{\Omega }\int_{T-\Omega }^{T}x'(s)ds}$  and involves the 
velocity of the evolution (regarded as a transaction in economic 
terms).  If one wishes to integrate in  the evaluation of 
deprivation   a  function  $\mathbf{l}: u \in X \mapsto  \mapsto 
\mathbf{l}(u) \in \mathbb{R}\cup \{+\infty \}$ of the 
transactions, we can add the  cumulated cost $\displaystyle{ 
\int_{T-\Omega }^{t} l(x'(s)) ds}$ of transactions $x'(s)$ for 
defining a new ``Willingness To Pay Valuation''.

Knowing both the evolutionary system $\mathcal{A}_{c}(T-\Omega 
,T;x)$,  the  cost function of the economic state at the 
beginning of the temporal window and the transaction cost 
function $\mathbf{l}$, one can define  a  ``Willingness To Pay 
Valuation'' $(T,\Omega ;x) \mapsto V(T,\Omega ;x)$ in the 
following way: 
\begin{equation} \label{e:WTPfct2}
W(T,\Omega ;x) \; := \; \inf_{x(\cdot)\in 
\mathcal{A}_{c}(T-\Omega ,T;x)} \left(
 \mathbf{c}(T-\Omega , x(T-\Omega )) + \int_{T-\Omega }^{T}\mathbf{l}(x'(t))dt \right)
\end{equation}
Optimal duration $\Omega _{\star}$ and evolutions 
$x_{\star}(\cdot) \in \mathcal{A}_{c}(T-\Omega ,T;x)$   
minimizing the Willingness To Pay Valuation valuation, if they 
exist, are regarded  as \emph{willingness to pay investment 
durations and evolutions}.

\section{The Value Function of an Optimal Control Problem} 
\label{ValueFct}

Let $X:=\mathbb{R}^{\ell}$ be a vector space (the commodity 
space). We denote by $x(\cdot):t \mapsto x(t) \in X$ a 
\emph{commodity evolution} (or ``flow''). Its derivative 
$x'(\cdot):t \mapsto x'(t) \in X$ is regarded as a 
\emph{transaction evolution}.

We introduce \index{temporal window} \emph{temporal window}  
$[T-\Omega ,T]$ where $\Omega \geq 0 $ is its \index{opening} 
\emph{opening} (and thus, $T-\Omega$ is the \index{departure 
time} \emph{departure time}).

\begin{Definition} 
\symbol{91}\textbf{Departure and Arrival 
Map}\symbol{93}\label{}\index{} We consider   two ``cost 
functions''  $\mathbf{c}$ and $\mathbf{l}$: 

\begin{enumerate}
\item an \index{instantaneous condition}\emph{instantaneous cost 
condition} function $ (t,x) \mapsto \mathbf{c}(t,x) \in 
\mathbb{R}\cup \{+\infty\} $;

\item a Lagrangian $\mathbf{l}: (t,x,u) \mapsto \mathbf{l}(t,x,u) 
\in \mathbb{R}\cup \{+\infty \}$, regarded as a \emph{transaction 
cost function} $u \mapsto \mathbf{l}(t,x,u)$ depending on time 
and commodity
 \end{enumerate}
with which we associate 
\begin{enumerate}   
\item the departure tube $C: \mathbb{R}^{}  \leadsto X$ defined by

\begin{displaymath}   
C(t) \; := \;  \left\{ x \in X \; \mbox{ such that} \; 
\mathbf{c}(t,x) \; <\; +\infty \right\}
\end{displaymath}   
\item  the set-valued map $F:\mathbb{R}^{} \times  X  \leadsto X$ 
defined by

\begin{equation} \label{e:}   
F(t,x) \; := \;  \left\{ u \in X \; \mbox{ such that} \; 
\mathbf{l}(t,x,u) \; <\; +\infty \right\}
\end{equation}
and the \index{arrival map} \emph{arrival map} 
 $\mathcal{A}_{\mathbf{l}}: \mathbb{R}^{} \times X  \leadsto \mathcal{C}(  
-\infty,+\infty;X) $  associating with any final pair $(T,x)$ the 
set of evolutions $x(\cdot)$ governed by the \index{differential 
inclusion} \emph{differential inclusion} 
\begin{equation} \label{e:LX-DifIncl}
\forall \;  t \in \mathbb{R}^{}, \; \;  x'(t) \; \in \;  
F(t,x(t)) 
\end{equation}
starting at some $s:= x(T-\Omega ) \in C(T-\Omega )$ and arriving 
at the prescribed terminal condition $x(T)   x$.
\end{enumerate}
 
\end{Definition}

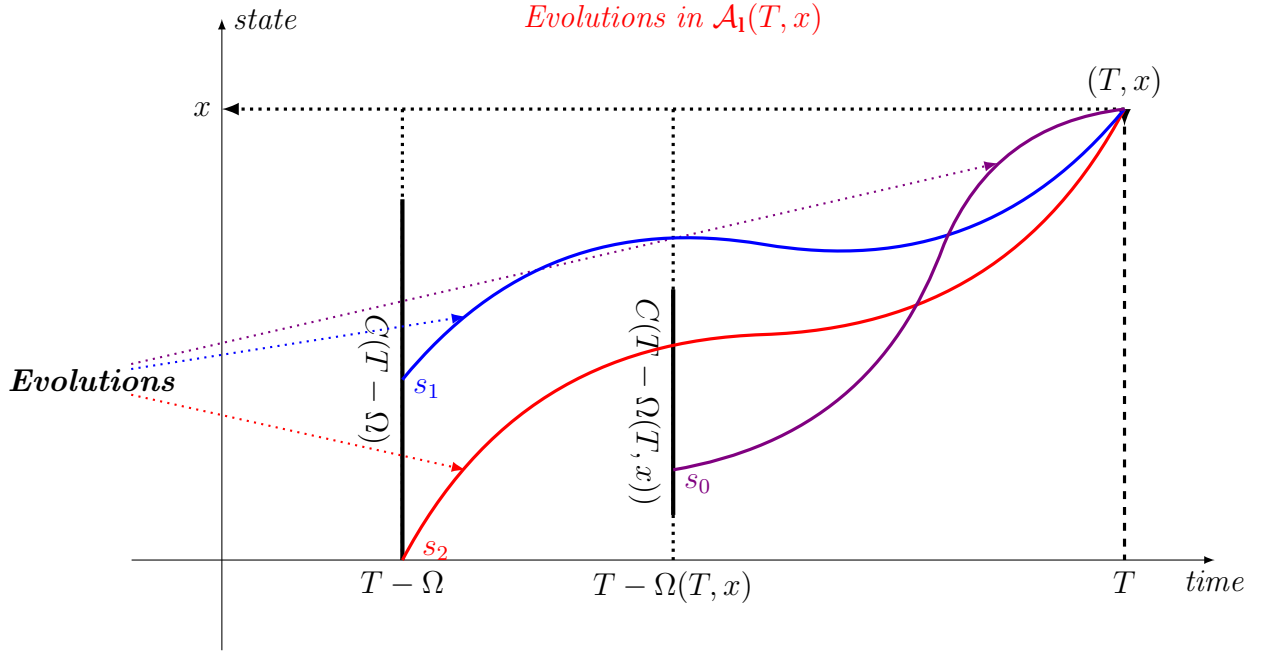
\begin{center}
\begin{tikzpicture}  [scale = 1.2]
\node[text width=7cm, red, text centered]  (Z) at (5,6 ) 
{\emph{Evolutions in \textbf{$\mathcal{A}_{\mathbf{l}}(T,x)$}}};  
 
\draw (10,5) node[above]{$(T,x)$}; 

\draw[->,>=latex, ] (-1,0) -- (11,0);

\draw[->,>=latex, ] (0, -1) -- (0,6);

\draw (10,0) node[below]{$T$};

\draw (0,5) node[left]{$x$}; 

\draw (2,0) node[below]{$T-\Omega $};

\draw (5,0) node[below]{$T-\Omega (T,x)$};

\draw (11,0) node[below]{\emph{time}};

\draw (0,6) node[right]{\emph{state}}; 

\draw[dotted, very thick] (2,5) -- (2,0); 

\draw[ultra thick] (2,4) -- (2,0) node[midway,below,sloped] {$ 
C(T-\Omega ) $}; 

\draw[dotted, very thick] (5,5) -- (5,0); 

\draw[ultra thick] (5,3) -- (5,.5) node[midway,below,sloped] {$ 
C(T-\Omega(T,x) ) $};     

\draw[->,>=latex, dotted, very thick] (10,5) -- (0,5); 

\draw[>-,>=latex, dashed, very thick] (10,5) -- (10,0);

\node[text width=1.5cm, text centered]  (B) at (-1.75,2) 
{\emph{\textbf{Evolutions}}}; 

\draw[->,>=latex,dotted, thick,  red] (B) -- (2.7,1); 

\draw[->,>=latex,dotted, thick, blue] (B) -- (2.7,2.7); 

\draw[->,>=latex,dotted, thick,  violet] (B) -- (8.6,4.4); 


\draw[very thick, red]   (2,0)  to[bend left]  (6,2.5) to[bend 
right]  (10,5);

\draw (2.1,0.1) node[red, right]{$s_{2}$};

\draw[very thick, blue]   (2,2)  to[bend left]  (6,3.5) to[bend 
right]  (10,5); 

\draw (2,1.9) node[blue, right]{$s_{1}$};

\draw[very thick, violet]   (5,1)  to[bend right]  (8,3.5) 
to[bend left]  (10,5); 

\draw (5,0.85) node[violet, right]{$s_{0}$};

\end{tikzpicture} \end{center}

In this study, we shall look for both 
\begin{enumerate}   
\item an \index{opening} \emph{opening} $\Omega \geq 0$  of the 
temporal window; 

\item  an \emph{evolution }$x(\cdot): [T-\Omega ,T] \mapsto X$ 
belonging to $\mathcal{A}_{\mathbf{l}}(T,x)$ (regulated by the 
differential inclusion $F$ and arriving at $x$ at time $T$) 

\end{enumerate}
satisfying the following optimality criterion:

\begin{Definition} 
\symbol{91}\textbf{Value Function}\symbol{93}\label{}\index{} The 
\emph{value function} of the associated intertemporal 
optimization problem with respect to the aperture $\Omega \geq 0$ 
and $x(\cdot) \in \mathcal{A}_{\mathbf{l}}(T,x)$ is defined by
 \begin{equation} \label{e:ValueFctChoux0} 
V(T,x) \;:= \; \inf_{\Omega \geq 0} \inf_{x(\cdot)  \in 
\mathcal{A}_{\mathbf{l}}(T,x)}\left(\mathbf{c}(T-\Omega 
,x(T-\Omega )) + \int_{T-\Omega }^{T}   \mathbf{l}(t, x(t), 
x'(t))dt\right) 
\end{equation} 
This optimization problem \emph{minimizes \begin{enumerate}   
\item the aperture $\Omega $ of the temporal window $[T-\Omega ,T 
]$;  \item   the sum of the initial condition at departure time 
$T-\Omega$ and the cumulated sum of the transaction costs on the 
temporal window $[T-\Omega,T]$.
\end{enumerate}}  

\end{Definition}

We refer to \emph{Time and Money. How Long and How Much Money is 
Needed to Regulate  a Viable  Economy},\cite[Aubin]{TM}, for the 
theorem stating the existence of optimal evolutions to this 
infinite dimensional problem as well as to Chapter~14, p. 563, of
\emph{Viability Theory.  New Directions}, \cite[Aubin, Bayen \&  
Saint-Pierre]{absp} for more examples. 

The purpose of this study is to adapt  to the case of general 
transaction-cost functions $\mathbf{l}: (t,x, u) \mapsto 
\mathbf{l}(t,x,u)$  the Lax-Hopf formula proved for convex 
transaction-cost functions $\mathbf{l}: u \mapsto \mathbf{l}(u)$ 
independent of $t$ and $x$.

\section{The Lax-Hopf formula} \label{LHf}

When the transaction cost function $u \mapsto \mathbf{l}(u)$ is 
convex and lower semicontinuous  (i.e., when its epigraph ${\cal 
E}p(\mathbf{l})$ is convex and closed) and \emph{depend neither 
on time nor on commodity}, the celebrated \emph{Lax-Hopf 
formula}\footnote{See \emph{Partial Differential Equations}, 
\cite[Evans]{Evans98}, \emph{Semiconcave Functions, 
Hamilton-Jacobi Equations, and Optimal Control}, \cite[Cannarsa 
\& Sinestrari]{CannarsaSinestrari}, \cite[Aubin, Bayen \& 
Saint-Pierre]{absp06hj}, \cite[Claudel \& Bayen]{CB010,CB011}, 
\cite[Désilles]{Desilles} and Section~11.5, p. 465, of 
\emph{Viability Theory.  New Directions}, \cite[Aubin, Bayen \&  
Saint-Pierre]{absp}.} states that the value function $V$ can be 
drastically simplified: 
\begin{equation} \label{e:Hopf}   
V(T,x) \; := \; \inf_{\Omega \geq 0}\inf_{\Upsilon \in \mbox{\rm 
Dom}(\mathbf{l})}\left(\mathbf{c}(T-\Omega ,x(T-\Omega )) + 
\Omega  \mathbf{l}(\Upsilon)\right)
\end{equation}
where $\Upsilon \in \mbox{\rm Dom}(\mathbf{l}) \subset X$ range 
over the domain\footnote{The domain $\mbox{\rm Dom}(\mathbf{l})$ 
is the subset of $\Upsilon \in X$ such that the transaction cost
$\mathbf{l}(\Upsilon) <+\infty $ is finite.} of the transaction 
cost function $\mathbf{l}$. 

Indeed, the infinite dimensional optimization problem 
(\ref{e:ValueFctChoux0}), p. \pageref{e:ValueFctChoux0} is 
reduced to a finite dimensional optimization (\ref{e:Hopf}), p. 
\pageref{e:Hopf} on $\mathbb{R}^{}_{+} \times X$.

It allows us to solve analytically the optimization problem and 
to simplify its numerical calculation. 

At optimal aperture $\Omega_{\star}$ and transaction 
$\Upsilon_{\star} \in \mbox{\rm Dom}(\mathbf{l})$,  evolutions 
$x_{\star}(\cdot) \in \mathcal{A}_{\mathbf{l}}(T,x)$ achieve the 
minimum of the value function:

\begin{equation} \label{e:HopfOpt}   
 \frac{V(T, 
x_{\star}(T))-V(T-\Omega_{\star},x_{\star}(T-\Omega_{\star})}{\Omega_{\star}}    
\; = \; \mathbf{l}\left( 
\frac{x_{\star}(T)-x_{\star}(T-\Omega_{\star})}{\Omega_{\star}}\right)
\end{equation}
which states that  \emph{the \index{enrichment} \emph{enrichment} 
of the optimal value function is equal to the transaction cost of 
the average transaction} on the temporal window 
$[T-\Omega_{\star}) ,T]$.

Furthermore, the dynamic optimality property 

\begin{equation} \label{e:}   
\forall \; t \in [T-\Omega_{\star},T], \; \;V(t) \;  = \; V(t, 
x_{\star}(t)) \; := \; 
\mathbf{c}(T-\Omega_{\star},x_{\star}(T-\Omega_{\star})) + 
\int_{T-\Omega_{\star}}^{t} \mathbf{l}(x_{\star}'(\tau)) d\tau
\end{equation}
holds true and satisfies the boundary conditions $V(T) = 
V(T,x)=V(T,x_{\star}(T))$ and  $V(T-\Omega_{\star}) =  
V(T-\Omega_{\star},x_{\star}(T-\Omega_{\star})) = 
\mathbf{c}(T-\Omega_{\star},x_{\star}(T-\Omega_{\star}))$.

\section{The Generalized Lax-Hopf Formula} \label{CLHf} 

The purpose of this study is to extend this Lax-Hopf  formula to 
the case when \emph{the transaction cost function depends on the 
time and/or the commodities}.  In this case, we introduce the 
concept of moderation of a transition cost function:

\begin{Definition} 
\symbol{91}\textbf{Moderation of a Transition Cost 
Function}\symbol{93} \label{d:ModLagr}\index{}  The 
\index{moderated transaction cost function} \emph{moderated 
transaction cost function} $(T,x,\Omega ,\Upsilon) \leadsto 
\Lambda_{\mathbf{l}} (T,x,\Omega ,\Upsilon)$ of a the transaction 
cost function $(t,x,u) \mapsto \mathbf{l}(t,x,u)$ is the value 
function of the intertemporal problem

\begin{equation} \label{e:}   
\Lambda_{\mathbf{l}} (T,x,\Omega ,\Upsilon) \; := \; 
\inf_{\{x(\cdot)\in \mathcal{A}_{\mathbf{l}}(T,x)|\frac{1}{\Omega 
} \int_{T-\Omega }^{T} x'(s)ds =  \Upsilon\}} \frac{1}{\Omega } 
\int_{T-\Omega }^{T}   \mathbf{l}(t, x(t), x'(t))dt 
\end{equation}
which depends on \begin{enumerate}   \item the aperture $\Omega 
\geq 0$ of the temporal window;  \item the (average) transactions 
$\Upsilon \in X$.  
\end{enumerate}
\end{Definition}

\textbf{Remark} --- \hspace{ 2 mm}  The function 
$\Lambda_{\mathbf{l}}$ does not depend upon the instantaneous 
cost function $\mathbf{c}$ and can be computed \textrm{off-line} 
for each pair $(\Omega ,\Upsilon)$. If it exists, the optimal 
evolutions $x_{(\Omega ,\Upsilon)}(\cdot) \in 
\mathcal{A}_{\mathbf{l}}(T,x) $ satisfying

\begin{equation} \label{e:}  
\frac{1}{\Omega }\int_{T-\Omega }^{T} x_{(\Omega 
,\Upsilon)}'(s)ds =  \Upsilon \; \mbox{and}\; 
\Lambda_{\mathbf{l}} (T,x,\Omega ,\Upsilon) \; := \;   
\frac{1}{\Omega } \int_{T-\Omega }^{T}   \mathbf{l}(t, x_{(\Omega 
,\Upsilon)}(t), x'_{(\Omega ,\Upsilon)}(t))dt 
 \end{equation} are computed once and for all.
\hfill $\;\; \blacksquare$ \vspace{ 5 mm}

The moderation of   a convex lower semicontinuous transition cost 
function coincides it when it depends only on transitions:

\begin{Lemma} 
\symbol{91}\textbf{Moderation of a Lower Semicontinuous Convex 
Function Depending Only on Transitions}\symbol{93} 
\label{l:ModCx}\index{} If the transition cost function 
$\mathbf{l}:u  \leadsto \mathbf{l}(u)$ in independent of time and 
commodity and convex and lower semicontinuous, then it coincides 
with its moderation $\Lambda_{\mathbf{l}}$:
\begin{equation} \label{e:}   
\forall \; \Upsilon \in \mbox{\rm Dom}(\mathbf{l}), \; \;
\Lambda_{\mathbf{l}} (T,x,\Omega ,\Upsilon) \; := \; 
\mathbf{l}(\Upsilon)
\end{equation}
\end{Lemma} 
\textbf{Proof} --- \hspace{ 2 mm} Since the evolution defined by 
$x_{\Upsilon}(t):= x - \Upsilon (T-t)$  belongs to 
$\mathcal{A}_{\mathbf{l}}(T,x)$, then 

\begin{equation} \label{e:}   
\Lambda_{\mathbf{l}} (T,x,\Omega ,\Upsilon)  \; \leq \;  
\frac{1}{\Omega } \int_{T-\Omega }^{T} 
\mathbf{l}(x_{\Upsilon}'(t))dt \; = \;  \mathbf{l}(\Upsilon)
\end{equation}
The opposite inequality follows from   the \index{Jensen 
inequality} the \emph{Jensen inequality} \glossary{Jensen (Johan) 
[L. W.]} (see \cite[Jensen]{Jensen}) stating that whenever 
$\mathbf{l}:u \mapsto \mathbf{l}(u)$ is lower semicontinuous and 
convex, then

\begin{equation} \label{e:}   
 \mathbf{l}(\Upsilon) \; = \; \mathbf{l} \left( \frac{1}{\Omega } \int_{T-\Omega }^{T} x'(s)ds 
\right) \leq  \frac{1}{\Omega } \int_{T-\Omega 
}^{T}\mathbf{l}(x'(s))ds     
\end{equation}
implying that $\mathbf{l}(\Upsilon) \leq \Lambda_{\mathbf{l}} 
(T,x,\Omega ,\Upsilon)$, so that  $\mathbf{l}(\Upsilon) = 
\Lambda_{\mathbf{l}} (T,x,\Omega ,\Upsilon)$. \hfill $\;\; 
\blacksquare$ \vspace{ 5 mm}

\begin{Theorem} 
\symbol{91}\textbf{The Generalized Lax-Hopf Formula for Commodity 
Dependent Transaction Costs}\symbol{93} \label{t:GenLH}  The  
\index{generalized Lax-Hopf formula} \emph{generalized Lax-Hopf 
formula} states that the value function $V$ defined by 
\begin{displaymath}
V(T,x) \;:= \; \inf_{\Omega \geq 0} \inf_{x(\cdot)  
}\left(\mathbf{c}(T-\Omega ,x(T-\Omega )) + \int_{T-\Omega 
}^{T}   \mathbf{l}(x'(t))dt\right) 
\end{displaymath} 
is equal to 
\begin{equation} \label{e:Luxi1}   
V(T,x) \; := \; \inf_{\Omega \geq 0}\inf_{\Upsilon \in 
X}\left(\mathbf{c}(T-\Omega ,x(T-\Omega )) + \Omega  
\Lambda_{\mathbf{l}} (T,x,\Omega ,\Upsilon)\right)
\end{equation}
which is a finite dimensional  minimization problem on 
$\mathbb{R}^{}_{+} \times X$. \\This is the generalization of the 
Lax-Hopf formula (\ref{e:HopfOpt}), p. \pageref{e:HopfOpt}  when 
the transition cost function  the average transition cost $u  
\leadsto \mathbf{l}(u)$ is convex, lower semicontinuous and  not 
depends only on transactions since  $\Lambda_{\mathbf{l}} (T,x; 
\Omega  , \Upsilon )=\mathbf{l}(\Upsilon)$ in this case.
\end{Theorem}
 
\textbf{Proof} --- \hspace{ 2 mm} The proof is as simple as the 
proof of Lax-Hopf formula.

\begin{enumerate}   \item   
We can write the value function $V$ in the form

\begin{equation} \label{e:ValueFctChoux1}  \left\{ \begin{array}{l} 
V(T,x) \;:= \; \inf_{\Omega \geq 0} \inf_{\Upsilon} 
\inf_{\{x(\cdot)  \in \mathcal{A}_{\mathbf{l}}(T,x) \; \mbox{ 
such that} \; 
\int_{T-\Omega }^{T} x'(t)= \Omega  \Upsilon\}}\\
\displaystyle{ \left(\mathbf{c}(T-\Omega ,x(T-\Omega )) + 
\int_{T-\Omega }^{T}   \mathbf{l}(t, x(t), x'(t))dt\right)} 
\end{array} \right. 
\end{equation}
which is equal to

\begin{equation} \label{e:ValueFctChoux1}  \left\{ \begin{array}{l} 
V(T,x) \;:= \; \inf_{\Omega \geq 0} \inf_{\Upsilon} 
 \left(\mathbf{c}(T-\Omega ,x(T-\Omega ))  \right. \\ 
\displaystyle{+ \left( \inf_{\{x(\cdot)  \in 
\mathcal{A}_{\mathbf{l}}(T,x) \; \mbox{ such that} \; 
\int_{T-\Omega }^{T} x'(t)= \Omega  \Upsilon\}}  \int_{T-\Omega 
}^{T}   \mathbf{l}(t, x(t), x'(t))dt\right)} 
\end{array} \right. 
\end{equation}
This is the formula which we were looking for;

\item  Lax-Hopf formula (\ref{e:HopfOpt}), p. \pageref{e:HopfOpt} 
follows from Lemma~\ref{l:ModCx}, p.\pageref{l:ModCx} and 
Theorem~\ref{t:GenLH}, p.\pageref{t:GenLH}. \hfill $\;\; 
\blacksquare$ \vspace{ 5 mm} \end{enumerate}

We recall the  if we assume that the cost function $\mathbf{l}$ 
is Marchaud,  the optimization problem has a solution.

\begin{Definition} 
\symbol{91}\textbf{Marchaud Transition Cost 
Functions}\symbol{93}\label{}\index{} \label{d:MarchaudFunctions} 
We shall say that a transaction cost function  $(t,x,u) \mapsto 
\mathbf{l}(t,x;u) \; \in \; \mathbb{R}\cup \{+\infty \}$ is 
\emph{Marchaud} if it is a lower semicontinuous function convex 
with respect to $u$ and if there exists a finite positive 
constants $c >0 $ such that
\begin{equation} \label{e:MarchaudLagrangian}
 \left\{ \begin{array}{l}
\mbox{\rm Dom}(\mathbf{l}(t,x; \cdot)) \; \subset \; c(\|x\| + \|d\|+1)B\;\mbox{\rm and is closed} \\
\forall \; u \; \in \; \mbox{\rm Dom}(\mathbf{l}(t,x; \cdot)),
 \; \; 0  \; \leq\; \mathbf{l}(d,x; u) \; \leq\; c(\|x\| + \|d\|+1)\\
\end{array} \right. \end{equation}
\end{Definition}

Under this condition, the value function inherits the properties 
of optimization problems: see Theorem~13.4.2, p. 533, 13.5.1, p. 
538 and 13.5.2, p. 539  of \emph{Viability Theory.  New 
Directions}, \cite[Aubin, Bayen \&  Saint-Pierre]{absp} that we 
summarize in the next statement.

\begin{Theorem} 
\symbol{91}\textbf{Existence and Properties of Optimal 
Evolutions}\symbol{93}\label{}\index{} Let us assume that the 
transaction-cost function $\mathbf{l}$ is a \emph{Marchaud 
function}  and that the initial cost function $\mathbf{c}$ is 
lower semicontinuous. Then there exist an optimal 
aperture and an optimal evolution.\\
At optimal aperture $\Omega_{\star}$ and optimal average 
transaction $\Upsilon_{\star}$, optimal evolutions 
$x_{\star}(\cdot)$ satisfy
\begin{equation} \label{e:}   
\Upsilon_{\star} \;= \; \frac{1}{\Omega_{\star} } 
\int_{T-\Omega_{\star} }^{T}x_{\star}'(t)dt \;\mbox{\rm and}\; 
\Lambda_{\mathbf{l}} (T,x,\Omega_{\star} ,\Upsilon_{\star}) \;  = 
\;   \frac{1}{\Omega_{\star} } \int_{T-\Omega_{\star} }^{T}   
\mathbf{l}(t, x_{\star}(t), x_{\star}'(t))dt
\end{equation}
and   the Isaacs-Bellman dynamic  optimal property stating that 
the  function $V: [T-\Omega ,T] \mapsto V(t)$ defined by
\begin{equation} \label{e:}   
V(t) \; := \; V(t, x_{\star}(t)) \; := \; 
\mathbf{c}(T-\Omega_{\star},x_{\star}(T-\Omega_{\star})) + 
\int_{T-\Omega_{\star}}^{t}  \mathbf{l}(t, x_{\star}(t), 
x_{\star}'(t))dt
\end{equation}
is still the optimal value function at $(t, x_{\star}(t))$  
satisfying  $V(T) = V(T,x)=V(T,x_{\star}(T))$ and  
$V(T-\Omega_{\star}) =  
V(T-\Omega_{\star},x_{\star}(T-\Omega_{\star})) = 
\mathbf{c}(T-\Omega_{\star},x_{\star}(T-\Omega_{\star}))$.\\
The generalized Lax-Hopf condition states that

\begin{equation} \label{e:Luxi2}  
\Lambda_{\mathbf{l}} \left( T,x,\Omega_{\star} 
,\frac{x_{\star}(T)-x_{\star}(T-\Omega_{\star})}{\Omega_{\star}}  
\right) \; = \; \frac{V(T, x_{\star}(T))-V(T-\Omega_{\star}, 
x_{\star}(T-\Omega_{\star}))}{\Omega _{\star}}
\end{equation} 
so that the \emph{enrichment 
 $\displaystyle{\frac{V(T)-V(T-\Omega_{\star} )}{\Omega_{\star}} }$  of the 
optimal value function $t \mapsto  V(t):= V(t,x_{\star}(t))$ on 
the temporal window $[T-\Omega_{\star} ,T]$ is equal the optimal 
moderated  transaction cost $\Lambda_{\mathbf{l}} 
(T,x,\Omega_{\star} ,\Upsilon_{\star})$. }
\end{Theorem}

\textbf{Example} --- \hspace{ 2 mm} In the case when 
$\mathbf{c}(t,x) =0$ when $t=0$ and $x=0$ and  $\mathbf{c}(t,x) 
=+\infty$ otherwise, the intertemporal optimal value function 
boils down to
\begin{equation} \label{e:ValueFctChoux1} 
V(T,x) \;:= \; \inf_{\Omega \geq 0} \inf_{\{x(\cdot)  \in 
\mathcal{A}_{\mathbf{l}}(T,x) \; \mbox{ such that} x(T-\Omega 
)=0\} }  \int_{T-\Omega }^{T}   \mathbf{l}(t, x(t), x'(t))dt  
\end{equation} 
The above formula (\ref{e:Luxi2}), p. \pageref{e:Luxi2} boils 
down to 
\begin{equation} \label{e:Luxi3} \left\{ \begin{array}{l}   
\displaystyle{\Upsilon_{\star} \;= \; 
\frac{x_{\star}(T)}{\Omega_{\star}} }
\\
\displaystyle{ \Lambda_{\mathbf{l}} (T,x; \Omega_{\star} , 
\Upsilon_{\star}) \; = \; \frac{V(T,x_{\star})}{\Omega_{\star}}  }
\end{array} \right. \hfill \;\; \blacksquare 
\end{equation}
 
\vspace{ 5 mm} 

\textbf{Remark} --- \hspace{ 2 mm} We  recall that the value 
function, when it is differentiable, is a solution to the 
\emph{Hamilton-Jacobi equation}. This was for solving 
Hamilton-Jacobi equations that the Lax-Hopf formula was derived. 
Nowadays, we can associate with any transaction function 
$\mathbf{l}  : (t,x,u) \mathbb{R}^{} \times X \times X \mapsto 
\mathbf{l}^{\star}(t,x,u) \times \mathbb{R}\cup \{+\infty \}$ its 
conjugate function\footnote{In physics, when $\mathbf{l}$ is 
interpreted as a Lagrangian, its conjugate function is called an 
Hamiltonian.} (also called the Legendre-Fenchel transform) 
$\mathbf{l}^{\star} : (t,x,p) \mathbb{R}^{} \times X \times 
X^{\star} \mapsto \mathbf{l}^{\star}(t,x,p) \times \mathbb{R}\cup 
\{+\infty \}$ defined by

\begin{equation} \label{e:}   
\mathbf{l}^{\star}(t,x,p) \; := \;  \sup_{u \in X} (\left\langle  
p,u\right\rangle - \mathbf{l}(t,x,u))
\end{equation}

The Fenchel theorem states that whenever $u  \leadsto 
\mathbf{l}(t,x,u)$ is convex and lower semicontinuous, then its  
conjugate $p  \leadsto \mathbf{l}^{\star}(t,x,p)$ is also convex 
and lower semicontinuous, and, furthermore, that the conjugate  
$\mathbf{l}^{\star^{\star}} =\mathbf{l}$ of $\mathbf{l}^{\star}$ 
is equal to $\mathbf{l}$.

Recall also that by definition, $p \in \partial 
\mathbf{l}(t,x,u)$ belongs to the subdifferential of $\mathbf{l}$ 
if and only if $\left\langle  p,u\right\rangle = 
\mathbf{l}(t,x,u)+ \mathbf{l}^{\star}(t,x,p)$, so that both are 
equivalent to  $u \in \partial \mathbf{l}(t,x,p)$.  The 
Hamilton-Jacobi equation associated with $\mathbf{l}^{\star}$ is

\begin{equation} \label{e:}   
\frac{\partial V(t,x)}{\partial t} \; := \;  \mathbf{l}^{\star} 
\left(t,x, \frac{\partial V(t,x)}{\partial x}\right) 
\end{equation}

The condition associated with the cost function $\mathbf{c}$ is 
written

\begin{equation} \label{e:}   
\forall \; (t,x), \; \; V(t,x) \; \leq  \;  \mathbf{c}(t,x)
\end{equation}  
 
When $\mathbf{l}$ is Marchaud, the value function, when it is 
differentiable, is a solution to the Hamilton-Jacobi equation.  
Otherwise, when it is not differentiable, but only lower 
semicontinuous, we can give a meaning to a solution as a 
solution  in the Barron-Jensen/Frankoska sense, using for that 
purpose subdifferential of lower semicontinuous functions defined 
in non-smooth analysis (\emph{Set-valued analysis}, \cite[Aubin 
\& Frankowska]{af90sva}, \cite[Aubin, Bayen, 
Saint-Pierre]{absp06hj,absp}). So, under this assumption, the 
value function if and only if it is a generalized solution of the 
Hamilton-Jacobi equation and if and only if it is a ``viability'' 
solution. So we can prove the Lax-Hopf formula for 
Hamilton-Jacobi equation in two steps, the first one uses the 
fact that the solution is the value function, the second one, bu 
using the generalized Lax-Hopf formula for the value function. 
The adaptation of the results of Chapters~13 and 17 of 
\emph{Viability Theory.  New Directions}, \cite[Aubin, Bayen \&  
Saint-Pierre]{absp}  is straightforward. \hfill $\;\; 
\blacksquare$ \vspace{ 5 mm}

\section{Lax-Hopf Formula with Transaction-Dependent Interest Rates} 
\label{IRCLHf}

We next introduce \emph{time, commodity and transaction  
dependent interest rate} $\mathbf{m}(t,x,u)$. Such interest rates 
can be prescribed constants $\mathbf{m}$, as it usually assumed, 
or prescribed time dependent rates $\mathbf{m}(t)$, or, more 
interestingly, $\mathbf{m}(t,x(t),x'(t))$ dependent also on the 
evolution of the commodity and the transaction.

For every evolution $x(\cdot) \in 
\mathcal{A}_{(\mathbf{l},\mathbf{m})}(T,x)$, we set
\begin{equation} \label{e:}  
M_{x(\cdot)}(t)  \; := \;  \int_{0}^{t}\mathbf{m}(\tau, x(\tau), 
x'(\tau))d\tau 
\end{equation}
and define the value function with interest rates by

\begin{equation} \label{e:ValueFctChoux01} \left\{ \begin{array}{l} 
V_{(\mathbf{l},\mathbf{m})}(T,x) \; := \; \inf_{\Omega \geq 0} 
\inf_{x(\cdot)  \in \mathcal{A}_{(\mathbf{l},\mathbf{m})}(t,x)} \\
\left(
 e^{M_{x(\cdot)}(\Omega)}\mathbf{c}(T-\Omega, x(T-\Omega)) +  
 \int_{T-\Omega }^{T} 
e^{M_{x(\cdot)}(T-  \tau)} (\mathbf{l}(\tau, x(\tau), 
x'(\tau))d\tau)    \right)
\end{array} \right. 
\end{equation}
 
We introduce the value function  

\begin{equation} \label{e:}   
\Lambda _{(\mathbf{l},\mathbf{m})} (T,x,\Omega ,\Upsilon) \; := 
\; \inf_{\{x(\cdot)\in \mathcal{A}_{(\mathbf{l},\mathbf{m})} 
(T,x)| \int_{T-\Omega }^{T} x'(s)ds = \Omega \Upsilon\}} 
\frac{1}{\Omega } \int_{T-\Omega }^{T} e^{M_{x(\cdot)}(T- t)}  
\mathbf{l}(t, x(t), x'(t))dt 
\end{equation}

Therefore,

\begin{Theorem} 
\symbol{91}\textbf{The Lax-Hopf Formula for Commodity Dependent 
Transaction Costs with Interest Rates}\symbol{93}\label{}\index{} 
The value function $V$ defined by (\ref{e:ValueFctChoux01}), p. 
\pageref{e:ValueFctChoux01} is equal to 

\begin{equation} \label{e:Luxi11}  
V_{(\mathbf{l},\mathbf{m})}(T,x) \; = \; \inf_{(\Omega ,\Upsilon) 
\in \mathbb{R}^{}_{+} \times X }  \left( e^{M_{x(\cdot)}(\Omega 
)}\mathbf{c}(T-\Omega, x-\Omega \Upsilon) + \Omega  \Lambda 
_{(\mathbf{l},\mathbf{m})} (T,x,\Omega ,\Upsilon) \right)
\end{equation}
which is a finite dimensional  minimization problem on 
$\mathbb{R}^{}_{+} \times X$ as the Lax-Hopf formula with time, 
commodity and transaction interest rates and costs.\\
At optimal aperture $\Omega_{\star}$ and optimal average 
$\Upsilon_{\star}$, optimal evolutions satisfy
\begin{equation} \label{e:Luxi21}   
 \Lambda_{(\mathbf{l}, \mathbf{m})} (T,x; 
\Omega_{\star} , \Upsilon_{\star}) \; :=   
\frac{1}{\Omega_{\star} } \int_{T-\Omega_{\star} }^{T}  
e^{M_{x_{(\Omega_{\star} ,\Upsilon_{\star})}(\cdot)}(T- t)}  
\mathbf{l}(t, x_{(\Omega_{\star} ,\Upsilon_{\star})}(t), 
x'_{(\Omega_{\star} ,\Upsilon_{\star})})(t)dt  
\end{equation}
and the Isaacs-Bellman dynamic  optimal property stating that 
the  function $V: [T-\Omega ,T] \mapsto V(t) $  defined by
\begin{equation} \label{e:}   
V(t) \;  \; = \; 
e^{M_{x_{\star}(\cdot)}(\Omega_{\star})}\mathbf{c}(T-\Omega_{\star}, 
x_{\star}(T-\Omega_{\star})) +  
 \int_{T-\Omega _{\star}}^{t} 
e^{M_{x_{\star}(\cdot)}(t-  \tau)} (\mathbf{l}(\tau, 
x_{\star}(\tau), x_{\star}'(\tau))d\tau)
\end{equation}
is still the optimal actualized (at the end of the temporal 
window value) function $V(t):= 
\;V_{(\mathbf{l},\mathbf{m})}(t,x_{\star}(t))$ at $(t, 
x_{\star}(t))$. The generalized Lax-Hopf formula states that at 
optimum, 
\begin{equation} \label{e:} 
\Lambda_{(\mathbf{l}, \mathbf{m})} (T,x; \Omega_{\star} , 
\Upsilon_{\star}) \; :=   \frac{1}{\Omega_{\star} } 
\int_{T-\Omega_{\star} }^{T} e^{M_{x_{(\Omega_{\star} 
,\Upsilon_{\star})}(\cdot)}(T- t)}  \mathbf{l}(t, 
x_{(\Omega_{\star} ,\Upsilon_{\star})}(t), x'_{(\Omega_{\star} 
,\Upsilon_{\star})})(t)dt
\end{equation}
is \emph{the ratio between the optimal actualized profit and the 
optimal aperture of the temporal window}, and thus, the 
\index{actualized enrichment} \emph{actualized enrichment}  (at 
terminal time $T$) of the actualized value function $V$.

\end{Theorem}

\section{Generalized Lax-Hopf Formula for a Dynamic Economy} 
\label{DECLHf}

The dual $X^{\star}:= \mathbb{R}^{\ell ^{\star}}$ of the 
commodity space is the space of \index{price} \emph{prices}  $p 
=(p^{h})_{1 \leq h \leq \ell} : x \mapsto \left\langle 
p,x\right\rangle \in \mathbb{R}^{}$ associating with any 
commodity $x \in X$  its   \emph{value} 
$\displaystyle{\left\langle p,x\right\rangle \; := \; \sum_{h 
=1}^{\ell} p^{h}x_{h}}$.   

The velocity $p'(t)$ at time $t$  of the evolution of a price 
$p(\cdot)$ is regarded as    the \index{price fluctuation} 
\emph{price fluctuations}. The \index{impact of price 
fluctuation} \emph{impact of price fluctuation} $\left\langle 
p'(t),x(t) \right\rangle$ on a commodity $x(t)$ is related to the 
concept of \index{inflation} \emph{inflation}\footnote{Inflation 
is measured as the impact of price fluctuations $\left\langle 
p'(t),b  \right\rangle$ on a numéraire or a \index{consumer price 
indexe} \emph{consumer price indexes} $b \in X$: 
$\displaystyle{\left\langle p'(t),x(t) \right\rangle= \left(
 \frac{\left\langle p'(t),x(t) \right\rangle}{\left\langle
p'(t),b  \right\rangle} \right)\left\langle p'(t),b 
\right\rangle}$. }.

We consider a set of $n$ (economic) agents. We denote by $X^{n}$ 
the set of allocations $x := (x_{i})_{i=1, \ldots,n}$ of 
commodities $x_{i} \in X$ among $n$ agents.  The 
\index{patrimonial value} \emph{patrimonial value}\footnote{This 
is the simplest example of an ``economic potential'' chosen for 
the sake od simplicity. In physics, the gradient of a 
\index{potential function} \emph{potential function} $U: x 
\mapsto U(x)$ is interpreted as a force: along an evolution $t 
\mapsto x(t)$, $\displaystyle{\frac{d}{dt}U(x(t))= \left\langle 
\frac{\partial U(x(t))}{\partial x},x'(t)\right\rangle}$ is a 
\index{power} \emph{power}. In economics, the variable $x$ is 
replaced by the allocation-price pair $(p,x)$ and \emph{the 
impetus plays the role of the mechanical power}.}  $t \mapsto 
U(x(t),p(t))$ along an evolution $t \mapsto (x(t),p(t))$ is 
defined by
\begin{equation} \label{e:}   
U(x(t),p(t)) \; := \;    \sum_{i=1}^{n} \left\langle 
p_{i}(t),x_{i} (t) \right\rangle 
\end{equation}  
and its \index{impetus} \emph{``impetus''} $t \mapsto 
E(x(t),p(t))$  by  Definition~7.1,1, p. 107, of \emph{Time and 
Money. How Long and How Much Money is Needed to Regulate  a 
Viable  Economy}, \cite[Aubin]{TM}:

\begin{equation} \label{e:}  E(x(t),p(t)) \; := \;
\frac{d}{dt}U(x(t),p(t)) \; = \;  \sum_{i=1}^{n}   \left( 
\left\langle p_{i}(t),x'_{i}(t)  \right\rangle  +\left\langle 
p'_{i}(t),x_{i}(t)  \right\rangle    \right)
\end{equation} 
    
An  \index{instantaneous cost function} \emph{instantaneous cost 
function} $(T,x,p)  \mapsto \mathbf{c}(T,x,p)$   of the state 
$(x,p) \in X^{n} \times X^{\star}$ at instant $T:=[T,T]$ (of zero 
aperture) is a function which associates the cost of $(x,p)$ at 
\emph{instant} $T$, \emph{regarded as a temporal window with zero 
aperture}. 

We shall assume once and for all that this instantaneous cost 
function is \index{lower semicontinuous} \emph{lower 
semicontinuous}.

\begin{Definition} 
\symbol{91}\textbf{Impetus Cost Function}\symbol{93} 
\label{d:CanonicalDynamics} The \index{impetus cost function} 
\emph{impetus cost function} described by   \emph{a priori} 
\begin{enumerate}   
\item  a convex lower semicontinuous function     $\mathbf{l}: E 
\in  \mathbb{R}^{} \mapsto \mathbf{l}(E) \in \mathbb{R}^{}_{+}$   
with which we associate  the \index{impetus cost function} 
\emph{impetus cost function} $t \mapsto \mathbf{l}(E(x(t),p(t)))$;

\item  dynamical behaviors  described by bounds $\gamma(\cdot):= 
(\gamma_{0}(\cdot),\gamma_{i}(\cdot)_{i=1, \ldots,n})$ where:
\begin{enumerate} 

\item  bound  $0\leq \gamma_{0}(t)<+\infty$ on the norm of the 
price fluctuations; 

\item  bound  $0\leq \gamma_{i}(t)<+\infty$ on the norm of the 
commodity transactions of the agents $i, \;i=1, \ldots,n$.

 \end{enumerate} 
We introduce  the \index{impetus cost function} \emph{(dynamical) 
impetus cost function}  defined by
\begin{equation} \label{e:} 
\begin{array}{l}
\mathbf{l}_{\gamma}(E(x(t),p(t))) := \\
 \\
\left\{ 
\begin{array}{ccl} \mathbf{l}(E(x(t),p(t))) & \mbox{if} & 
\max(\max_{i}(\|x'_{i}(t)\| -c_{i}(t)),\|p'(t)\|- c_{0}(t)) \leq 0
\\
\\  + \infty & \mbox{if not} &\\
\end{array} \right. \end{array} 
\end{equation}
\end{enumerate}
 
We denote by 
$\mathcal{A}(T,x,p):=\mathcal{A}_{\mathbf{l}_{\gamma}}(T, x,p)$ 
the set of evolutions $(x(\cdot),p(\cdot)) $ of  evolutions   $t 
\mapsto x(t):=(x_{i}(t))_{i=1, \ldots,n} $ and $t \mapsto p(t) $ 
with bounded velocities arriving at $(x,p)$ at terminal time $T$.
\end{Definition}

We modify the concept of ``endowment function'' introduced in 
Chapter~7,p. 105, of \emph{Time and Money. How Long and How Much 
Money is Needed to Regulate  a Viable  Economy} 
,\cite[Aubin]{TM}), as suggested in Footnote~3, p.109 by 
introducing impetus cost functions.

\begin{Definition} 
\symbol{91}\textbf{The Economic Value 
Function}\symbol{93}\label{}\index{d:TMEndowmentFunction2} The 
instantaneous cost $\mathbf{c}$, the bounds of the canonical 
dynamic system   being given, the endowment function $W:(T,x,p) 
\mapsto (T,x,p)$ is defined by

\begin{equation} \label{e:} \left\{ \begin{array}{l }
W(T,x,p) \; := \; \inf_{\Omega \geq 0} \inf_{(x(\cdot),p(\cdot)) 
\in \mathcal{A} _{ \mathbf{l}_{\gamma}}(T,x,p)} 
\\\displaystyle{ \left( 
\mathbf{c}(T-\Omega,x(T-\Omega),p(T-\Omega)) + \int_{T-\Omega 
}^{T} \mathbf{l}_{\gamma}  (E(x(t),p(t)))dt  
 \right) }\\
\end{array} \right.
\end{equation}
which is the infimum over the set of  evolutions 
$(x(\cdot),p(\cdot))$ of their  cumulated impetus cost.
\end{Definition}

Since the impetus cost function depends upon the time, the 
commodity and the price, we have to introduce the moderated 
impetus cost function $\Lambda _{\mathbf{l}}(T,x,t, \Omega 
,\Upsilon_{i}, \Upsilon_{0})$  defined by

\begin{equation} \label{e:}   
\Lambda _{\mathbf{l}}(T,x,t, \Omega ,\Upsilon_{x}, \Upsilon_{p}) 
\; := \;  \inf_{\{\frac{1}{\Omega } \int_{T-\Omega }^{T} x' 
(t)=\Upsilon_{x} \Omega \}} \inf_{ \frac{1}{\Omega } 
\int_{T-\Omega }^{T} p'(t)=\Upsilon_{p} \Omega \} }  
\frac{1}{\Omega } \int_{T-\Omega }^{T} 
\mathbf{l}_{\gamma}(E(x(t),p(t)))dt
\end{equation}

Therefore,

\begin{equation} \label{e:}   
W(T,x,p) \; := \; \inf_{\Omega \geq 0} 
\inf_{\Upsilon_{x},\Upsilon_{p}} (\mathbf{c}(T-\Omega, x-\Omega 
\Upsilon_{x}, p- \Omega \Upsilon_{p}) + \Lambda 
_{\mathbf{l}}(T,x,t, \Omega ,\Upsilon_{x}, \Upsilon_{p}) )
\end{equation} 
Then, at optimum, the enrichment of the economic value function 
is equal to the moderated impetus function at average optimal 
transaction and price fluctuation on the temporal window:

\begin{equation} \label{e:} \left\{ \begin{array}{l}  
\displaystyle{\frac{W(T,x_{\star}(T),p_{\star}(T))-W(T-\Omega_{\star},x_{\star}(T-\Omega_{\star}),p_{\star}(T-\Omega_{\star})))}{\Omega 
_{\star}} }\\ \displaystyle{:= \;   \Lambda _{\mathbf{l}} \left( 
T,x,p, \frac{x_{\star}(T)-x_{\star}(T-\Omega _{\star})}{\Omega 
_{\star}} ,\frac{p_{\star}(T)-p_{\star}(T-\Omega 
_{\star})}{\Omega _{\star}}  \right) }
\end{array} \right. 
\end{equation}

\clearpage \tableofcontents
\end{document}